\documentstyle [12pt,axodraw] {article}

\catcode`@=12
\begin{document}
\thispagestyle{empty}
\begin{flushright} DESY 98-133\\ August 1998\
\end{flushright}
\vspace{0.5in}
\begin{center}
{\Large \bf $CP$ violation in the mass matrix of heavy neutrinos \\}
\vspace{0.7in}
{\bf Utpal Sarkar$^{1,2}$ and Rishikesh Vaidya$^{1}$\\}
\vspace{0.2in}
{$^1$ \sl Physical Research Laboratory, Ahmedabad 380 009, India\\}
\vspace{0.1in}
{$^2$ \sl Theory Group, DESY, D-22603 Hamburg, Germany\\}
\vspace{0.7in}
\end{center}
\begin{abstract}

We discuss the question of $CP-$violation in the effective Hamiltonian
approach in models of leptogenesis
through heavy right handed neutrino decays. We first formulate the
problem in four component notation and then point out that before the
heavy neutrinos have decayed away, the universe becomes $CP-$asymmetric 
in the heavy neutrinos. However, the lepton asymmetry
generated after they completely decay are independent of this asymmetry.

\end{abstract}

\newpage
\baselineskip 18pt

The baryon asymmetry of the universe could be created from a
lepton asymmetry of the universe before the electroweak phase 
transition \cite{fy,ma,luty}. In these scenarios the lepton number
violation at some very high energy generates a lepton asymmetry of 
the universe, which then get converted to a baryon asymmetry when 
the sphaleron mediated processes are in equilibrium. A popular 
scenario is the one in which the right handed neutrinos ($N_{R i}, i=
e, \mu, \tau$) decay into
light left-handed leptons and anti-leptons \cite{fy}. 

$CP-$asymmetries are calculated in the heavy neutrinos decay from
the interference of tree level and one-loop vertex and self energy
corrections. The self-energy \cite{ls} was initially
considered as one additional contribution to $CP$ asymmetry, 
but later it was pointed
out that this may be interpreted as an oscillation type of indirect
$CP$ violation \cite{paschos1}. Since then several works contributed
towards the understanding of this effect \cite{covi,paschos2,ris,small} 
and a resonant phenomenon was observed \cite{paschos2,small} for nearly
degenerate right handed neutrinos. 

The self-energy contribution to the $CP-$violation has been interpreted
as $CP-$violation in the mass matrix. This gives a difference between
the amplitudes of $N_{Ri} \to N^c_{Rj}$ and $N^c_{Ri} \to N_{Rj}$. 
In analogy to the $K-$system, we call this $CP-$violation 
as the indirect $CP-$violation
and the vertex contribution to the $CP-$violation in the decays of the
right handed neutrinos as the direct $CP-$violation. 
In earlier treatments of this effect, the neutrinos and 
antineutrinos were treated as different two component objects
\cite{paschos1,paschos2}, 
although finally the physical states were identified with four
component physical Majorana states. We formulate this problem in 
the conventional method of treating a Majorana particle as a
four component object, which will clarify several features of
the problem. We then discuss the
question of $CP-$violation for Majorana particles and point out
that a $CP-$asymmetric universe in the heavy neutrinos is created first, 
which is different from the $CP-$asymmetry 
created after these heavy neutrinos decay. We point out the essential
role played by the non-hermiticity of the Hamiltonian in this problem,
which was not appreciated before.

To simplify the problem we work in a two generation framework and
include two right handed neutrinos ($N_i$,  $i=1,2$)
and write down their Majorana mass term and their interactions with
the light fields,

\vbox{
\begin{eqnarray}
{\cal L}_{int} & = & \sum_{i} M_i [\overline{(N_{Ri})^c} N_{Ri} + 
   \overline{N_{Ri}} (N_{Ri})^c]
  + \sum_{\alpha, i} \, h^\ast_{\alpha i} \, \overline{N_{Ri}} 
 \, \phi^{\dagger} \,
     \ell_{L \, \alpha} + \sum_{\alpha, i} h_{\alpha i} \, 
      \overline{\ell_{L \, \alpha}} \phi \, N_{Ri} \nonumber \\ 
 & & + \sum_{\alpha, i} \, h^\ast_{\alpha i} \, \overline{(\ell_{L 
 \, \alpha}})^c \, \phi^\dagger \, (N_{Ri})^c + \sum_{\alpha, i} 
 h_{\alpha i} \overline{(N_{Ri})^c} \, 
       \phi \, (\ell_{L \, \alpha})^c  
\end{eqnarray}}

\noindent where  $\phi \equiv (1,2,1/2)$ is the usual higgs doublet
of the standard model and $\ell_{\alpha L}, \alpha = e, \mu, \tau$ are
the light left-handed leptons; $h_{\alpha i}$ are the Yukawa  couplings
and in general complex. We work in a basis  in  which  the
Majorana mass matrix is real and diagonal with eigenvalues $M_i$.

Once the phases of $N_i$ are fixed by the choice of the basis, we cannot
absorb any phases contained in the couplings $h_{\alpha i}$. A phase
transformation of the light leptons can apparently absorb some of these 
phasses, but they will then appear in the charged current interactions. 
So, if there is $CP-$violation in the leptonic sector, then phase
transformations of the light leptons cannot absorb any of the remaining
phases. In addition, measurable quantities should not depend
on the choice of phase of the light leptons. So to understand the
question of $CP-$violation, it is convenient to consider rephasing 
invariant quantities \cite{pal,acker}, 
which are the combinations of the couplings which remains
invariant under any change of phases of the light leptons. 

In the quark sector there is only one $CP-$phase in the CKM-matrix $V_{aj}$
and one can construct only one rephasing invariant quantity, which is the
Jarlskog invariant $J = V_{ai} V_{bj} V^*_{aj} V^*_{bi}$. 
In the leptonic sector \cite{pal} the Majorana
nature of the neutrinos allow new types of $CP-$invariant quantities, 
which may not have any analogy in the quark sector. Consider the phase
transformation of the light leptons,
$$ l_\alpha \to e^{i \delta_\alpha} l_\alpha $$ 
which will imply a phase tranformation to the Yukawa couplings,
$$ h_{\alpha i} \to e^{i \delta_\alpha} h_{\alpha i}.$$ 
Under this phase transformation the combinations of the Yukawa
couplings which remains invariant and could be complex are,
\begin{equation}
t_{\alpha i j} = {\rm Im} \:\: (h_{\alpha i} h^*_{\alpha j}) \label{taij}
\end{equation}
and
\begin{equation} 
T_{\alpha \beta i j} = {\rm Im} \:\: (h_{\alpha i} h_{\beta j} 
h^*_{\alpha j} h^*_{\beta i})  . \label{tabij}
\end{equation} 
Thus if any of the above combination of the Yukawa couplings enter
in some process, there can be $CP-$violation. 

The quantity $T_{\alpha \beta i j}$ is similar to the Jarlskog invariant
in the quark sector. However, there is no analog of the
other rephasing invariant quantity $t_{\alpha i j}$ in the quark sector
since there are no Majorana particles. In the decays of the right
handed neutrinos, the direct and the indirect $CP-$asymmetry depends on
the same rephasing invariant quantity ${\rm Im}~[t_{\alpha i j} 
t_{\beta i j}]$ \cite{acker}. In our present formalism we shall 
demonstrate that first we have a $CP-$asymmetric universe in $N_{Ri}$ 
and the amount of $N_{Ri}$ asymmetry depends on the rephasing invariant 
quantity ${\rm Im}~[t_{\alpha i j}]$. As a result it is possible to 
imagine a situation with ${\rm Re}~[t_{\alpha i j}] = 0$, but with 
${\rm Im}~[t_{\alpha i j}] \neq 0$, when the 
universe could be $CP-$asymmetric in $N_{Ri}$ at higher temperature, 
but after the heavy neutrinos decay the universe becomes $CP-$symmetric. 

We shall now discuss another conceptually important question of how
the $N_{Ri}$ asymmetry is generated before they decayed away. For Majorana
particles we donot have to consider the $N_{Ri}$ and
$N^c_{Ri}$ as independent, since they are related by Majorana condition.
The mass matrices are also same for both. However, when there is 
$CP-$violation, the mass matrices are no longer the same, but still they
remain related by simple phase rotation. 

The situation changes when these particles have a decay width. This point
has been discussed in details in the field theoretic language extensively
\cite{ris}, but here we shall discuss this point in an effective
Hamiltonian language. Consider the
tree level mass matrix in the $[N_{1}~~N_{2}]$ basis,
\begin{equation}
\left( \begin{array}{cc}
M_{1}&0\\
0&M_{2}
\end{array}\right)
\end{equation}
Since $CP$ is conserved at this level, the mass matrix is same as above in
the $[N_{1}^{c}~~N_{2}^{c}]$ basis, so that hermitian conjugate states may
be obtained by operating with CP on the physical states.
 
We can then write down the total effective Hamiltonian
including the one loop self energy corrections,
in the bases $[N_{1} N_{2}]$ and $[N_{1}^{c}N_{2}^{c}]$ respectively as,
\begin{equation}
{\cal H}^{+}=\left(\begin{array}{cc}
M_{1}&h^{+}\\
h^{+}&M_{2}
\end{array}\right)\hspace{.4in}
{\rm and} \hspace{.4in} {\cal H }^{-}=\left(\begin{array}{cc}
M_{1}&h^{-}\\
h^{-}&M_{2}
\end{array}\right) 
\end{equation}
where we have absorbed the dispersive part in the wave function and
mass renormalisation; redefined the diagonal elements as
$M_{a}\rightarrow ~M_{a} + {h_a}$; with,
\begin{equation}
h^+ = - {i \over 32 \pi}
\left[ M_i \sum_\alpha h_{\alpha i}^\ast h_{\alpha j} + 
 M_j \sum_\alpha h_{\alpha i} h_{\alpha j}^\ast \right]
\end{equation}    
\begin{equation}
h^- = - {i \over 32 \pi}
\left[M_i  \sum_{\alpha} h_{\alpha i} h_{\alpha j}^\ast
     + M_j \sum_{\alpha} h_{\alpha i}^\ast h_{\alpha j} \right] 
    \end{equation}
and 
\begin{equation}
h_a  = - {i \over 32 \pi}
\left[2 M_i \sum_{\alpha}  h_{\alpha i} h_{\alpha i}^\ast  \right] 
\end{equation}
neglecting terms of order $O \left( {m_\alpha^2}/{p^2} \right) $, 
$O \left( {m_\phi^2}/{p^2} \right) $ with $p^2 \geq M_i^2$.

Because of the absorbtive part the mass matrices of the $N_{Ri}$ and
$N^c_{Ri}$ are no longer related by just phase rotation. This needs some
explanation. Because of the decay width of the heavy neutrinos $N_{Ri}$, 
their mass matrix takes the form $M - {i \over 2} \Gamma$, where the 
decay width $\Gamma$ comes from the the absorbtive part and hence it
is anti-hermitian. As a result, the mass matrices of the charge
conjugate states $N^c_{Ri}$ now becomes $M^\ast - {i \over 2} \Gamma^\ast$
which is not related by simple phase rotation to the mass matrix of the
states $N_{Ri}$. 

It was not emphasized earlier that this anti-hermitian decay width 
$\Gamma$ implicitly gives $CPT$ violation at the {\it effective theory 
level}, and hence the conventional $CPT$ conserving results do not hold.
Consider the theorem which tells us that in thermal equilibrium 
the number density of particles with non-zero charge $Q$ 
would be same as the number density of the antiparticles since 
the expectation value of the conserved charge, given by,
$$ <Q> = \frac{{\rm Tr} \left[ Q e^{-\beta H}\right]}{{\rm Tr} \left[ 
e^{-\beta H}\right]} 
$$
vanishes since any conserved charge $Q$ is odd while $H$ is even 
under ${ CPT}$. Consider now a non-hermitian Hamiltonian, 
$$ H = M \overline{\Psi^c} \Psi $$
where $\Psi$ carries a charge $Q = 1$ and there is no hermitian 
conjugate part of the Hamiltonian. This Hamiltonian now satisfies,
$[Q,H] = 2 H$. It is now obvious that this charge $Q$ cannot be odd
or $H$ cannot be even under $CPT$, because then this commutation 
relation is not invariant under $CPT$. On the other hand, if $Q$
is not odd or $H$ is not even, then the above theorem does not hold
and one can have a non-zero expectation value for the charge $Q$.
Because of this reason, when the decay width is included in the
mass matrix along with its $CP$ non-conservation, an asymmetry
was obtained for $N_{Ri}$. 

It is similar to the fact, that the total decay width of any particle
is same as that of the total decay width of the antiparticle. But
the partial widths for them could be different. In fact, the partial
decay widths are just equal and opposite to each other for the 
different decay modes of particles and antiparticles. 
In the same way, if we consider the total
system, $CPT$ will be  conserved and we cannot get any asymmetry.
As a result, the lepton asymmetry can be
generated in this system only when the $N_{Ri}$ decays out-of-equilibrium. 

As an illustration of the discussions of the previous paragraph 
consider a scalar $X$, which has two decay modes,
$$ X \to a + b \hskip 1in X \to c + d .$$ 
$CPT$ invariance would then imply that the total decay widths satisfy, 
$\Gamma_{total} (X) = \Gamma_{total} (\bar{X})$. Whereas, $CP$ violation 
would imply that the partial decay widths of $X$ and $\bar{X}$ are different, 
$\Gamma (X \to a + b) \neq \Gamma (\bar{X} \to \bar{a} + \bar{b})$,
although $\Gamma (X \to a + b) = \Gamma (\bar{X} \to \bar{c} + \bar{d})$.
In analogy, when $CPT$ and unitarity is respected, we have 
\begin{equation}
\sum_f \Gamma (S_i \to S_f) = \sum_f \Gamma (S_f \to S_i), 
\label{uni}
\end{equation}
where $S_i$ and $S_f$ are the initial and final states. However, when
the system departs from equilibrium, this is no longer true locally.
Consider the case, when the two real intermediate states
$X_S$ and $X_L$ have very short and long lifetimes, which are combinations
of the states $X$ and $\bar{X}$. In a short interval of time, only $X_S$
will have time to decay, while $X_L$ will decay after the universe has
expanded further. So at any given time equation (\ref{uni}) is not valid,
which has been discussed in details in ref \cite{ris}.

In words, if $N_{Ri}$ decays and inverse decays are slow enough to
satisfy the out-of-equilibrium condition, the difference in the decay
widths of the physical states can generate an asymmetry in $N_{Ri}$. 
This is because although the state with fast decay rate can recombine
again, the slow decaying particle cannot recombine. These rates for the
particles and antiparticles are different because of the non-hermiticity 
of the decay matrix $\Gamma$. However, if the physical states decay
very fast, then the decays and inverse decays of these states will be
in equilibrium and we have to treat the system as a whole, which 
conserve $CPT$ and hence there will not be any $N_{Ri}$ asymmetry.

We shall now calculate the amount of $CP-$asymmetry in $N_{Ri}$
in this approach. In the 
present effective Hamiltonian language, the mass matrices of $N_{Ri}$ and
$N^c_{Ri}$ being different implies that due to the different 
Majorana masses the transition $N_{Ri} \to N^c_{Rj}$ (which is same as
$N_{Rj} \to N^c_{Ri}$ by the $CPT$ theorem) is not the same as that
of $N^c_{Rj} \to N_{Ri}$ (which is same as that of
$N^c_{Ri} \to N_{Rj}$ by $CPT$). This produces an $N-$asymmetric
universe even before their decay generates an asymmetry 
in the left-handed leptons. 

The eigenvalues for these Hamiltonians are
\begin{equation}
\lambda_{1}^{\pm}=\frac{1}{2}(M_{1}+M_{2}+\sqrt{S^{\pm}})
\hspace{.2in}and\hspace{.2in}
\lambda_{2}^{\pm}=\frac{1}{2}(M_{1}+M_{2}-\sqrt{S^{\pm}})
\end{equation}
where
\begin{equation}
 S_{\pm}=(M_{1}-M_{2})^{2}+4{h^{\pm}}^{2}. 
\end{equation}
The corresponding physical states are,\\
\begin{equation}
\Psi_{\stackrel{1}{2}}^{+}=a_{\stackrel{1}{2}}^{+}N_{1} +
b_{\stackrel{1}{2}}^{+}N_{2}
\end{equation}
\begin{equation}
\Psi_{\stackrel{1}{2}}^{-}=a_{\stackrel{1}{2}}^{-}N_{1}^{c} +
b_{\stackrel{1}{2}}^{-}N_{2}^{c}
\end{equation}
with,
\begin{equation}
\left(\begin{array}{cc}
a_{1}^{\pm}&b_{1}^{\pm}\\
a_{2}^{\pm}&b_{2}^{\pm}
\end{array}\right)
=\left(\begin{array}{cc}
\frac{{\cal C}^{\pm}}{{\cal N}^{\pm}}&\frac{1}{{\cal N}_{\pm}}\\
\frac{1}{{\cal N}_{\pm}}&-\frac{{\cal C}^{\pm}}{{\cal N}^{\pm}}
\end{array}\right).
\end{equation}
and
\begin{equation}
{\cal C}^{\pm}=\frac{2h^{\pm}}{M_{2}-M_{1}+\sqrt{{\cal S}^{\pm}}};
\hspace{.3in} and\hspace{.3in} {\cal N}^{\pm}=\sqrt{1+{({\cal C}^{\pm})}^2}.
\end{equation}

The physical states $\Psi_{1,2}^{\pm}$ will now evolve with time. So
even if we start with an $N$ symmetric universe, we will end up in a 
$N$ asymmetric universe. The physical states will decay and recombine
continuously. However, since the decay widths for the two physical
states are different, this process will create an asymmetry in $N_{Ri}$,
which will be compensated by an asymmetry in the decay products. If 
one calculates the amount of asymmetry in $N_{Ri}$ and the decay products,
the $CPT$ theorem will tell us that there cannot be any net asymmetry.
However, when we calculate the asymmetry of only $N_{Ri}$ when the
decay rates are slow enough to satisfy the out-of-equilibrium condition,
there will be an 
asymmetry in $N_{Ri}$. The amount of $N_{Ri}$ asymmetry is given by,
\begin{equation}
\Delta_N = \sum_{i=1,2} \frac{\Gamma_{\psi_{i}^{+}\rightarrow N}-
\Gamma_{\psi_{i}^{-}\rightarrow N^{c}}}
{\Gamma_{\psi_{i}^{+}\rightarrow N}+\Gamma_{\psi_{i}^{-}\rightarrow
N^{c}}} = {A - B \over A + B},
\end{equation}
where, $A = |a_i^+ + b_i^+|^2 = 2 |{\cal N}^+|^{-2} [ |{\cal C}^+|^2 +1]$ 
and $B = |a_i^- + b_i^-|^2 = 2 |{\cal N}^-|^{-2} [ |{\cal C}^-|^2 +1]$.

In the approximation, $h^\pm \ll |M_2 - M_1|$ we can find an expression
for the $N$ asymmetry to be,
\begin{equation}
\Delta_N = {1 \over 8 \pi} {M_1 - M_2 \over M_1 + M_2}
{\rm Im}~ \sum_\alpha h_{\alpha i}^\ast h_{\alpha j}
\end{equation}
where the rephasing invariant $CP$ violating quantity is 
of the type ${\rm Im} [t_{\alpha ij}]$.

In the present approach we can calculate the final lepton asymmetry
in the same way. The physical states $\psi^+_i$ will decay into
only leptons, whereas the states $\psi^-_i$ will decay into only
antileptons. Since the mass matrices for the states $N_{Ri}$ and
$N^c_{Ri}$ are not related by just phase rotation, there will be 
an asymmetry. In the field 
theoretic language this has been clarified and 
demonstrated recently \cite{ris}.

The amount of lepton number asymmetry is given by,
\begin{equation}
\delta=\sum_{i=1,2}\frac{\Gamma_{\psi_{i}^{+}\rightarrow l}-
\Gamma_{\psi_{i}^{-}\rightarrow l^{c}}}
{\Gamma_{\psi_{i}^{+}\rightarrow l}+\Gamma_{\psi_{i}^{-}\rightarrow
l^{c}}},
\end{equation}
where,
\begin{equation}
\Gamma_{\psi_{i}^{+}\rightarrow l}=\sum_{\alpha}|a_{i}^{+}h_{\alpha
1}+b_{i}^{+}h_{\alpha 2}|^2
\end{equation}
and
\begin{equation}
\Gamma_{\psi_{i}^{-}\rightarrow l^{c}}=\sum_{\alpha}|a_{i}^{-}h_{\alpha
1}^{*}+b_{i}^{-}h_{\alpha 2}^{*}|^2
\end{equation}
and hence,
\begin{equation}
\Delta=\sum_{i}\frac{{\cal X}^{i}_{-}}{{\cal X}^{i}_{+}}
\end{equation}
where
\begin{eqnarray}
{\cal X}^{i}_{\pm}&=&(|a_{i}^{+}|^2 \pm
|a_{i}^{-}|^{2})\sum_{\alpha}|h_{\alpha 1}|^{2}+(|b_{i}^{+}|^{2}\pm
|b_{i}^{-}|^{2})\sum_{\alpha}|h_{\alpha
2}|^{2} \nonumber \\
&&+2~Re[\sum_{\alpha}h_{\alpha
1}^{*}h_{\alpha 2}(a_{i}^{+*}b_{i}^{+}\pm a_{i}^{-}b_{i}^{-*})]
\end{eqnarray}
In the approximation $|M_2 - M_1| \gg |h^\pm|$, 
it is given by the expression as obtained by other methods, 
\begin{equation}
\delta = -{1 \over 8 \pi}{M_1 M_2 \over M_2^2 - M_1^2} 
{{\rm Im}~ \left[ \sum_\alpha (h_{\alpha 1}^\ast h_{\alpha 2})
 \sum_\beta (h_{\beta 1}^\ast h_{\beta 2}) \right] \over
 \sum_{\alpha} |h_{\alpha 1}|^2}
\end{equation}
where the rephasing invariant $CP$ violating phase is of the form
${\rm Im}~[t_{\alpha 1 2} t_{\beta 1 2}]$. As a result, it is 
possible to create a $N_{Ri}$ asymmetry, without creating a lepton
asymmetry of the universe. However, before generating lepton asymmetry
of the universe through heavy Majorana neutrino decay, we cannot avoid
having a $CP-$asymmetric universe in the right handed neutrinos. When
a $CP-$asymmetric universe is created in $N_{Ri}$, an equal and 
opposite amount of light left-handed lepton asymmetry is also generated
to compensate. So, with the same assignment of the lepton numbers
for $N_{Ri}$ and the light leptons, the total lepton asymmetry of 
the universe before the decay of the heavy neutrinos would vanish
because of the conservation of $CPT$. 
On the other hand, after the heavy neutrinos 
decay the generated asymmetry will depend on the amount of $CP-$violation
and more crucially depend on the departure of the system from 
thermal equilibrium. 

The Boltzmann equation has been solved
for the same system in details in the literature \cite{ris}, and
it has been pointed out that when the heavy right handed neutrinos
decay away from equilibrium, the $CP$ asymmetry produces a lepton
asymmetry of the universe. Because of a difference in the decay width
of the two physical states, when proper weight is given to the
different real intermediate states, the amount of $CP$ asymmetry
we calculated will produce a lepton asymmetry of the universe. 
During the electroweak phase transition, this lepton asymmetry 
will get converted to a baryon asymmetry of the universe. 

To summarize, we have studied the question of lepton asymmetry of
the universe in heavy neutrinos decay. We have developeded the
effective hamiltonian formalism with four component neutrinos for
this problem. We point out that before generating a lepton
asymmetry, the universe becomes $CP-$asymmetric in the heavy 
neutrinos even before the inverse decay stops. We then calculate the
amount of lepton asymmetry finally created after the heavy neutrinos
decayed away completely, which is different from the  $CP$ asymmetry 
of the heavy neutrinos. 

\noindent {\bf Acknowledgement} \\ One of us (US) would like to thank
Prof. W. Buchmuller and the Theory Division of DESY for hospitality
and acknowledge financial support from the Alexander von Humboldt 
Foundation.


\newpage
\begin{thebibliography}{99}

\bibitem{fy} M. Fukugita and T. Yanagida, Phys. Lett. {\bf B 174}
  (1986) 45.

\bibitem{ma} E. Ma and U. Sarkar, Phys. Rev. Lett. {\bf 80} (1998) 5716.

\bibitem{luty} P.  Langacker, R.D.  Peccei and T.  Yanagida, Mod.
  Phys.  Lett.  {\bf  A  1}  (1986)  541;  M.A.  Luty, Phys.
  Rev.  {\bf D 45}  (1992)  455;  R.N.  Mohapatra  and X.  Zhang,
  Phys.  Rev.  {\bf D 46} (1992) 5331; A.  Acker, H.  Kikuchi, E.
  Ma and U.  Sarkar,  Phys.  Rev.  {\bf D 48}  (1993)  5006;
  P.J. O'Donnell and U.  Sarkar,  Phys.  Rev.  {\bf
  D 49} (1994) 2118; W. Buchmuller and M. Pl\"{u}macher, Phys.
  Lett. {\bf B 389}, 73 (1996); A. Ganguly, J.C. Parikh
  and U. Sarkar, Phys. Lett. {\bf B 385} (1996) 175;
  M. Pl\"{u}macher, Z. Phys. {\bf C 74}, 549 (1997);
  
\bibitem{ls} J. Liu and G. Segre, Phys. Rev. {\bf D 48} (1993) 4609.

\bibitem{paschos1} M. Flanz, E.A. Paschos and U. Sarkar, Phys. Lett.
  {\bf B 345} (1995) 248.

\bibitem{covi}  L. Covi, E. Roulet and F. Vissani, Phys. Lett.
  {\bf B 384}, 169 (1996).

\bibitem{paschos2} M. Flanz, E.A. Paschos, U.  Sarkar
  and J. Weiss, Phys. Lett. {\bf B 389}, 693 (1996).

\bibitem{ris}  L. Covi, E. Roulet and F. Vissani, Phys. Lett.
  {\bf B 424} (1998) 101;
  W. Buchm\"{u}ller and M. Pl\"{u}macher, hep-ph/9710460;
  M. Flanz and E.A. Paschos, hep-ph/9805427.

\bibitem{small}  A. Pilaftsis, Phys. Rev. {\bf D 56} (1997) 5431;
  L. Covi and E. Roulet, Phys. Lett. {\bf B 399} (1997) 113;

\bibitem{pal} J.F. Nieves and P.B. Pal, Phys. Rev.
  {\bf D 36}, 315 (1987).

\bibitem{acker} A. Acker, H. Kikuchi, E. Ma and 
  U. Sarkar, Phys. Rev. {\bf D 48} (1993) 2118.

\end{thebibliography}
\end{document}